\documentclass[aps,pra,showpacs,twocolumn]{revtex4} 

\usepackage[english]{babel}

\usepackage{latexsym}
\usepackage{graphicx}
\usepackage{subfigure}
\usepackage{epsfig}
\usepackage{amsfonts}
\usepackage{amssymb}
\usepackage{amsmath}
\usepackage{bm, bbm}
\usepackage{mathbbol}

\begin{document}

\title{A view on the open STIRAP problem}

\author{Thomas Mathisen}
\affiliation{Department of Physics,
Stockholm University, AlbaNova University Center, 106 91 Stockholm,
Sweden}

\author{Jonas Larson}
\affiliation{Department of Physics,
Stockholm University, AlbaNova University Center, 106 91 Stockholm,
Sweden}

\date{\today}

\begin{abstract}
We consider the open STIRAP problem by formulating it in terms of generalized Bloch equations. Expressed in this form the resulting Liouvillian matrix is analyzed in detail. As it turns out, most of the findings of open STIRAP models can be understood from the spectrum and eigenstates of the Liouvillian matrix. As a non-hermitian matrix we particularly discuss the mathematical structure of it by identifying among others the physical meaning of its eigenvectors and the system's exceptional points. We also discuss the importance of the Liouvillian gap which normally implies that any state becomes mixed. However, in certain situations, for example the STIRAP model under influence of spontaneous emission, a sort of pump term appears in the equations of motion, and its effect is to counteract the gap induced decay such that coherence may prevail. As the STIRAP driving is not perfectly adiabatic, we additionally find that for spontaneous emission the presence of a non-zero Liouvillian gap actually enhances the success rate of the process. This is thus an example of a environment assisted physical operation. 
\end{abstract}
  
\maketitle

\section{Introduction}
{\it Coherent control} has become an essential part of many branches in quantum physics, ranging from atomic and molecular thermal gases to solid state devises and ultracold atomic condensates~\cite{shore,cc}. Utilizing adiabatic driving is a method to circumvent any errors arising from timing of applied pulses -- as long as the process is adiabatic the target state is reached. One of the most famous, and also the simplest, example is that of a {\it Landau-Zener} sweep~\cite{cclz}, where population is transferred between two quantum states via adiabatic following through an avoided crossing. While insensitive to precise timing, this method is, however, prone to spontaneous emission of the excited state. This can be bypassed by using instead the so called {\it STIRAP} -- stimulated Raman adiabatic passage method~\cite{shore,vitanovrev}. The idea of STIRAP is not to couple the two states directly but via an intermediate, possibly excited, state using two pulses. By properly choosing the coupling pulses, essentially their order, coherent transfer of population between the two lower (stable) states can be made perfect in the adiabatic limit, without ever populating the excited intermediate state. As such, deficiencies due to spontaneous emission of that state are greatly suppressed~\cite{sponem1,sponem2}. However, dephasing of the two lower states may well occur, for example due to elastic collisions between particles~\cite{dephase1}. Thus, the process is no longer coherent and this indeed affects the population transfer.    
 
In AMO experiments, the coupling to any environment can usually be made rather weak and especially when  working in the optical regime the system dynamics can be well approximated by a {\it Markovian master equation}. The general form of such an equation is given by the {\it Lindblad equation}~\cite{open}
\begin{equation}\label{master}
\partial_t\hat\rho = \mathcal{L}[\hat\rho]=i\left[\hat\rho,\hat H\right]+\mathcal{D}[\hat\rho],\
\end{equation}
with
\begin{equation}
\mathcal{D}[\hat\rho]=\displaystyle{\sum_i\gamma_i\left(2\hat L_i\hat\rho\hat L_i^\dagger-\hat L_i^\dagger\hat L_i\hat\rho-\hat\rho\hat L_i^\dagger\hat L_i\right).}
\end{equation}
Here, the first part represents the unitary time evolution governed by the system Hamiltonian $\hat H$, and the second part incorporates the effects of the environment. The effective system-environment couplings are $\gamma_i$ with corresponding {\it Lindblad jump operators} $\hat L_i$ that together with the Hamiltonian characterize the type of dissipation/decoherence acting upon the system. Applications of the Lindblad master equation to the open STIRAP problem, both for spontaneous emission and dephasing, have been done in the past~\cite{sponem2,dephase1}. This was done by either adiabatic approximate methods or direct numerical simulations. As anticipated, it was found that in the adiabatic regime spontaneous emission could indeed be suppressed, while dephasing remains crucial at all time-scales. 

In the present work we analyze the same problem, also with the Lindblad master equation as a starting point, but with a new approach. We parametrize the density operator in terms of generalized {\it Bloch vectors} and express 
Eq.~(\ref{master}) in terms of these. The master equation is then recast into a standard first order matrix differential equation, with a non-Hermitian ({\it Liouvillian}) matrix generating the time-evolution. The equation does not necessarily become homogeneous. Expressed in this form we analyze both the case of spontaneous emission and dephasing. The eigenvalues of the Liouvillian matrix (all with non-positive real parts) give direct information about the system evolution. Despite the finite size of the matrix, its eigenvalues/eigenvectors may become non-analytic in so called {\it exceptional poins} (EP's)~\cite{expo}. At these points, two (complex) eigenvalues merge and their corresponding eigenvectors become identical and the Liouvillian matrix is then non-diagonalizable. The eigenvalue with the smallest amplitude of the real parts defines the {\it Liouvillian gap} which determines the time-scale for the system to reach steady state. Since the Liouvillian matrix is explicitly time-dependent, the instantaneous gap should be compared to the time-scale for the STIRAP. In particular, if the STIRAP process is long the environment will have a greater influence on the system dynamics, and possibly deteriorate the success rate. Simultaneously, a too fast STIRAP will imply non-adiabatic excitation taking you out from the instantaneous eigenstate. As a result we find an optimal time-scale for the STIRAP process in the presence of dephasing. The situation is, as already mentioned, different for spontaneous emission where a slow process is always favourable. Shortly, the instantaneous adiabatic eigenstate that the system follows is also a so called {\it dark state} for the Lindblad jump operators meaning that it is transparent to the environment. Nevertheless, when the process is not perfectly adiabatic we actually find that the environment has the effect to increase the population transfer to the target state. This is understood, once again, from the Liouvillian gap that tends to project the state back onto the desirable dark state. 

The paper is structured as follows. In the next section we start by recapitulating the general idea behind STIRAP, and then move on to the open STIRAP problem where we summarize some earlier results. The following subsection discusses general properties of the Liouvillian matrix. We point out some remarks that seem overlooked by many, for example that in general an eigenstate of the Liouvillian does not represent a physical state, only a few of them do. Having build up some intuition for the master equation given in the Bloch representation we continue in Sec.~\ref{sec3} with the actual numerical results. In particular, the spectrum of the Liouvillian matrix is analyzed in some detail. And the full open STIRAP problem is simulated numerically, which confirms our predictions drawn from the structure of the Liouvillian matrix. Finally, we conclude in Sec.~\ref{sec4} with a summary and some remarks.

\section{The open STIRAP model} 

\subsection{STIRAP for closed systems}
\begin{figure}[h]
\centerline{\includegraphics[width=6cm]{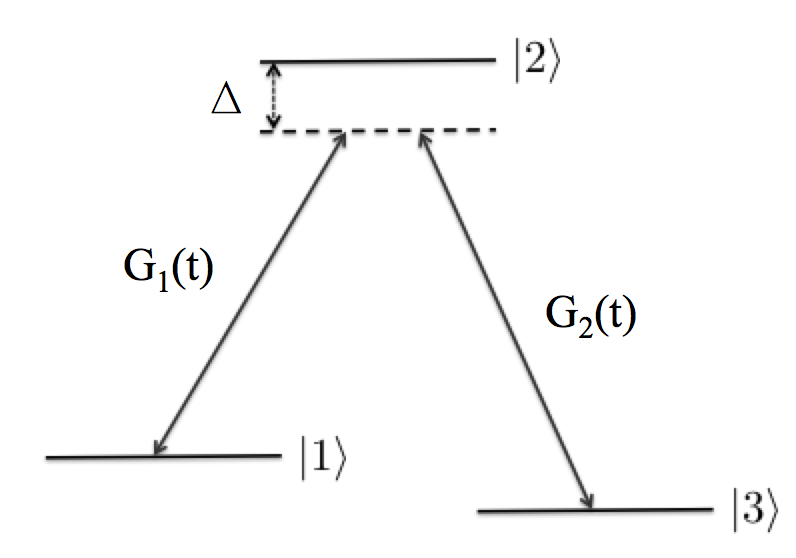}}
\caption{The traditional STIRAP $\Lambda$-scheme. Two pulses couple the lower stable states $|1\rangle$ and $|3\rangle$ with the excited state $|2\rangle$ with strengths $G_1(t)$ and $G_2(t)$. The lengths of the arrows symbolize the frequencies (in scaled units) of the light pulses, such that $\Delta$ marks the detuning between the applied pulses and the atomic transitions. In the figure the two-photon transition is resonant such that only a single detuning parameter is included.} \label{fig1}
\end{figure}

The standard STIRAP setup is the $\Lambda$ one depicted in Fig.~\ref{fig1}; Two stable states, $|1\rangle$ and $|3\rangle$ respectively, are laser coupled by a {\it pump} $G_1(t)$ and a {\it Stokes field} $G_2(t)$ to an excited intermediate state $|2\rangle$. Throughout we assume a two-photon resonance transition, {\it i.e.} the photon frequency difference $\hbar(\omega_2-\omega_1)$ (with $\omega_{1,2}$ the frequency of the pump and Stoke lasers respectively) matches the energy difference between the bare states $|1\rangle$ and $|3\rangle$. For the main part of this analytical subsection we do not restrict the analysis to single photon resonance transitions, and thereby the introduction of a detuning $\Delta$~\cite{shore}. It is found, however, that the qualitative results will not depend on $\Delta$ and for all numerical simulations we thereby let $\Delta=0$ for simplicity.

Within the {\it rotating wave approximation} and using the bare ({\it diabatic}) basis introduced above, the Schr\"odinger equation becomes~\cite{shore} ($\hbar=1$ throughout)
\begin{equation}\label{stirapham}
i\partial_t|\Psi(t)\rangle=\hat{H}_\mathrm{d}(t)|\Psi(t)\rangle=\!\left[
\begin{array}{ccc}
0 & G_1(t) & 0 \\
G_1(t) & \Delta & G_2(t) \\
0 & G_2(t) & 0
\end{array}\right]\!|\Psi(t)\rangle.
\end{equation}
This equation defines the time-dependent Hamiltonian $\hat H_\mathrm{d}$, and furthermore we have assumed real couplings $G_{1,2}(t)$ without loss of generality. The {\it adiabatic basis} $\left\{|\varphi_+(t)\rangle,\,|\varphi_0(t)\rangle,\,|\varphi_-(t)\rangle\right\}$ is given by the instantaneous (adiabatic) eigenstates of $\hat{H}_\mathrm{d}(t)$. Written as a unitary matrix given in the diabatic basis, these states are
\begin{equation}\label{adstate}
\hat{U}(t)=\left[
\begin{array}{ccc}
\sin\phi\sin\theta & \cos\theta & \cos\phi\sin\theta\\
\cos\phi & 0 & -\sin\theta\\
\sin\phi\cos\theta & -\sin\theta & \cos\phi\cos\theta
\end{array}\right],
\end{equation}
with the parametrization $\tan\theta=G_1(t)/G_2(t)$, $\tan2\phi=2G_0(t)/\Delta$, and with $G_0^2(t)=G_1^2(t)+G_2^2(t)$. The corresponding instantaneous (adiabatic) eigenstates are
$E_\pm(t)=\left(\Delta\pm\sqrt{\Delta^2+4G_0^2(t)}\right)/2$ and $E_0(t)=0$. With these we naturally have $\hat{H}_\mathrm{ad}(t)\equiv\hat{U}\hat{H}_\mathrm{d}(t)\hat{U}^{-1}=\mathrm{diag}\left(E_+(t),\,E_0(t),\,E_-(t)\right)$, with $\hat{H}_\mathrm{ad}(t)$ the {\it adiabatic Hamiltonian}. The Schr\"odinger equation in the adiabatic basis reads
\begin{equation}
i\partial_t|\Psi_\mathrm{ad}(t)\rangle=\left[\hat{H}_\mathrm{ad}(t)-i\hat{U}^{-1}\partial_t\hat{U}\right]|\Psi_\mathrm{ad}(t)\rangle.
\end{equation}
The {\it adiabatic approximation} consists in dropping the `gauge potential' $\hat{A}=-i\hat{U}\partial_t\hat{U}^{-1}$ that comprises the non-adiabatic couplings~\cite{shore,BOA,adtheo}. Thus, within this approximation, the adiabatic states evolve as $|\varphi_i(t)\rangle \rightarrow \exp\left(-i\int_0^tE_i(t')dt'\right)|\varphi_i(t)\rangle$ with $i=\pm,\,0$~\footnote{Here we put $\hat A\equiv0$, but it is noted that this gauge potential is responsible for the Berry phase when encircling a closed loop in parameter space.}. The {\it dark state} $|\varphi_0(t)\rangle$ is particularly attracting for practical purposes as it does not contain the bare excited state $|2\rangle$ which is typically subject to spontaneous emission. Now, if the couplings are chosen 
\begin{equation}\label{pulses}
\begin{array}{c}
\displaystyle{\lim_{t\rightarrow-\infty}\frac{G_1(t)}{G_2(t)}=0,\hspace{0.7cm}\theta\rightarrow0},\\ \\ 
\displaystyle{\lim_{t\rightarrow+\infty}\frac{G_2(t)}{G_1(t)}=0,\hspace{0.7cm}\theta\rightarrow\frac{\pi}{2}}
\end{array}
\end{equation}
it follows that, provided that the evolution is adiabatic, the dark state obeys
\begin{equation}
|\varphi_0(t)\rangle=\left\{
\begin{array}{lll}
|1\rangle, & \hspace{0.5cm} & t=-\infty,\\ \\
|3\rangle, & \hspace{0.5cm} & t=+\infty.
\end{array}\right.
\end{equation}
This defines the STIRAP in our $\Lambda$-configuration; if we prepare our state in $|1\rangle$ and adiabatically turn on the couplings according to (\ref{pulses}) we will steer the state into $|3\rangle$ without ever populating the mediating state $|2\rangle$. One simple choice of couplings are two symmetric Gaussians
\begin{equation}
\begin{array}{c}
\displaystyle{G_1(t)=g_0\exp\left[-\frac{(t-a\tau)^2}{2(a\sigma)^2}\right]},\\ \\
\displaystyle{G_2(t)=g_0\exp\left[-\frac{(t+a\tau)^2}{2(a\sigma)^2}\right]}
\end{array}
\end{equation}
Here, $g_0$ is the pulse amplitude, $a\sigma$ the pulse width, and $2a\tau$ the distance between the pulses. These are the pulses used throughout this manuscript. What is especially worth pointing out is the counterintuitive order of the pulses; pulse 2, the {\it pump}, which couple the initially empty states is turned on before pulse 1, the {\it Stokes}.  We have introduced $a$ to serve as our single `adiabaticity parameter' (qualitatively it is the pulse area setting the level of adiabaticity~\cite{vitanovrev}, and hence one could imagine varying other parameters, like the pulse amplitude, instead). Thus, we will keep $g_0$, $\tau$ and $\sigma$ fixed in all numerical simulations (more precisely $g_0=1$ and $\tau=\sigma=10$) and instead vary $a$ alone. To get a better insight into the parameters rendering adiabatic evolution we make use of the criteria for adiabaticity~\cite{adtheo}
\begin{equation}\label{adcrit}
\frac{\left|\langle\varphi_+(t)|\left(\partial_t\hat{H}_\mathrm{d}\right)|\varphi_0(t)\rangle\right|}{E_+^2(t)}\ll1,
\end{equation}   
where we have used the symmetry of the $|\varphi_\pm(t)\rangle$ adiabatic states relative to $|\varphi_0(t)\rangle$, and that $E_0(t)=0$. This results in
\begin{equation}\label{cond1}
\mathcal{A}(t)\equiv\frac{G_1(t)G_2(t)}{G_0^3(t)}\frac{a\tau}{(a\sigma)^2}\ll1.
\end{equation}
For adiabatic evolution, condition (\ref{cond1}) should be fulfilled for all times $t\in[t_i,t_f]$ between initial and final times. For not too large initial and final times, $\mathcal A(t)$ peaks at $t=0$, and we directly notice that since $\mathcal A(t)\sim1/g_0$ and $\mathcal A(t)\sim1/a$ a large amplitude $g_0$ and/or a large $a$ favour adiabaticity. As we already pointed out, we fix $g_0=1$ and instead vary $a$ in order to analyze the influence of non-adiabatic excitations. As we will see, for the open STIRAP problem, $a$ is in general not a good adiabaticity parameter. 

\begin{figure}[h]
\centerline{\includegraphics[width=8cm]{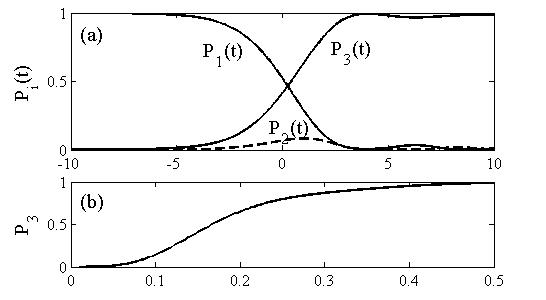}}
\caption{Bare/diabatic state populations $P_i$ ($i=1,\,2,\,3$) as a function of time (upper plot (a) with $a=1/2$) and the adiabaticity parameter $a$ (lower plot (b)). The integration is over the interval $t\in[-100,+100]$ which is well beyond the convergence of the populations. The remaining dimensionless parameters are $\Delta=0$, $\sigma=\tau=10$ and $g_0=1$. Note, in particular, that the population of the intermediate state $|2\rangle$ is non-vanishing during the transition period, even though it is almost fully absent at the final time. Thus, the final time fidelity $P_3$ is not a complete measure of how well the state followed its corresponding adiabatic (dark) state. Nevertheless, taking $P_3$ as a measure of adiabaticity it is seen that $a$ acts as a proper adiabaticity parameter as $P_3$ is a monotonously increasing function of $a$. } \label{fig2}
\end{figure} 

An example of typical evolution of the populations
\begin{equation}
P_i(t)=\langle i|\hat{\rho}(t)|i\rangle,\hspace{0.7cm}i=1,\,2,\,3,
\end{equation}
with $\hat{\rho}(t)=|\Psi(t)\rangle\langle\Psi(t)|$ the density operator, is shown in Fig.~\ref{fig2} (a). Here we also define $P_i=\lim_{t\rightarrow+\infty}P_i(t)$. As is seen, the population is adiabatically swapped between the two states $|1\rangle$ and $|3\rangle$. The transfer is, however, not perfect for this choice of parameters. In the (b) plot we depict how the final population of the state $|3\rangle$ depends on the parameter $a$. Expectedly, for short effective pulse durations the process is {\it diabatic} and very little population is transfered to the $|3\rangle$ state, while for $a>0.5$ the process is predominantly adiabatic. 

\subsection{The open STIRAP master equation}
There are in particular two types of dissipation/decoherence processes occurring naturally in realistic settings, spontaneous emission of the excited state $|2\rangle$ and dephasing. It has been shown that spontaneous emission plays a minor role on the STIRAP efficiency as long as the excited state is negligibly populated~\cite{sponem1,sponem2}. On the other hand, dephasing may severely affect the population transfer~\cite{dephase1}.

One of the first studies on how spontaneous emission influences the STIRAP consisted in adding a complex component $-i\gamma|2\rangle\langle2|$ to the Hamiltonian~\cite{sponem1}. In this simplified picture it was found, via adiabatic elimination, that the transition probability $P_i$ for $i=3$ falls off exponentially in terms of the dissipation rate $\gamma$ for weak dissipation $g_0\gg\gamma$, and as $\gamma^{-2}$ in the opposite limit $\gamma\gg g_0$. Naturally, in this simple model the Hamiltonian is not hermitian, and in particular, it does not take fluctuations deriving from the environment into account. The decay of the amplitude of the state can be seen as if population is lost from the excited state $|2\rangle$ to some auxiliary states lying outside our three-level manifold. If the lower states $|1\rangle$ and $|3\rangle$ are the only accessible lower sattes the population loss of $|2\rangle$ has to end up in these. Reference~\cite{sponem2} approaches this scenario by considering the full Lindblad master equation. As such it includes not only dissipation of the excited state, but also decoherence arising from the coupling to the environment. Again, after an adiabatic elimination of the intermediate state analytical expressions for the final populations are derived. In the limit of large decay and an intermediate change of the Hamiltonian one finds no population transfer as expected; the short time-scale of the problem is the one of the decay so population cannot leave the initial state, {\it i.e. } {\it quantum overdamping}. Taking the limit of vanishing decay, the analytical approximations recovers the closed system result of perfect transfer in the adiabatic limit. For a large regime of decay rates and being deep in the adiabatic regime, their results confirmed that the success rate was in principle unaltered by the openness of the problem.

In a more recent work~\cite{stirapmaster}, the open STIRAP was reexamined by deriving a microscopic master equation which take into account the explicit time-dependence of the Hamiltonian~\footnote{In the standard derivation of  the Lindblad master equation, the Hamiltonian is assumed time-independent. The derivation cannot be performed following the same procedure once the Hamiltonian becomes time-dependent. }. Thus, the decay channels are not added phenomenologically as in previous studies. The resulting Lindblad master equation differs from the phenomenological ones as the jump operators are constructed from the adiabatic states. A similar problem also arises in light-matter systems when the coupling is in the {\it deep ultrastrong regime}; the relevant process here is not decay of photons but rather decay of polaritonic states which are the eigenstates of the Hamiltonian~\cite{strong}. In the overdamped regime it was found that an increased population transfer is to be found, which results from a Zeno-type process. However, it should be noted that in order for this to be seen, the decay rate is by far the shortest time-scale and the imposed approximations of the model, {\it secular}, {\it Born}, and {\it Markov}, may all break down in this regime.  

Dephasing between the two states $|1\rangle$ and $|3\rangle$ is expected to have a greater effect on the STIRAP process. Dissipation of the excited state can be argued to have a small influence as long as this state is negligibly populated. Dephasing among the lower states will, on the other hand, effectively couple the dark state to the two other adiabatic `{\it bright}' states, and hence result in deterioration of the population transfer success rate. Influence of dephasing on the STIRAP problem has been analyzed in Ref.~\cite{dephase1} in the weak system-environment coupling limit. In this regime it is legitimate to assume that the interaction time, $T_\mathrm{int}\sim a\sigma$, is short compare to the time-scale for dephasing, $T_2\sim\gamma^{-1}$ (with $\gamma$ the effective system-environment coupling). In such a case, as shown in~\cite{dephase1}, using adiabatic elimination schemes, the final state population of the target state $|3\rangle$ can be estimated
\begin{equation}
P_3=\frac{1}{3}+\frac{2}{3}e^{-3\gamma aT^2/4\tau}.
\end{equation}

In the present work we reexamine the open STIRAP problem for the two cases mentioned above. We do so by using a slightly different approach, namely by rewriting the Lindblad master equation as a linear differential equation for the Bloch vector of the state. In two dimensions the resulting equations are the famous {\it Bloch equations}~\cite{be}. In higher dimension, however, much less is known about these `generalized Bloch equations', and the interpretation and analysis of the equations become also much more complex as will be discussed in detail below. 

\begin{figure}[h]
\centerline{\includegraphics[width=8cm]{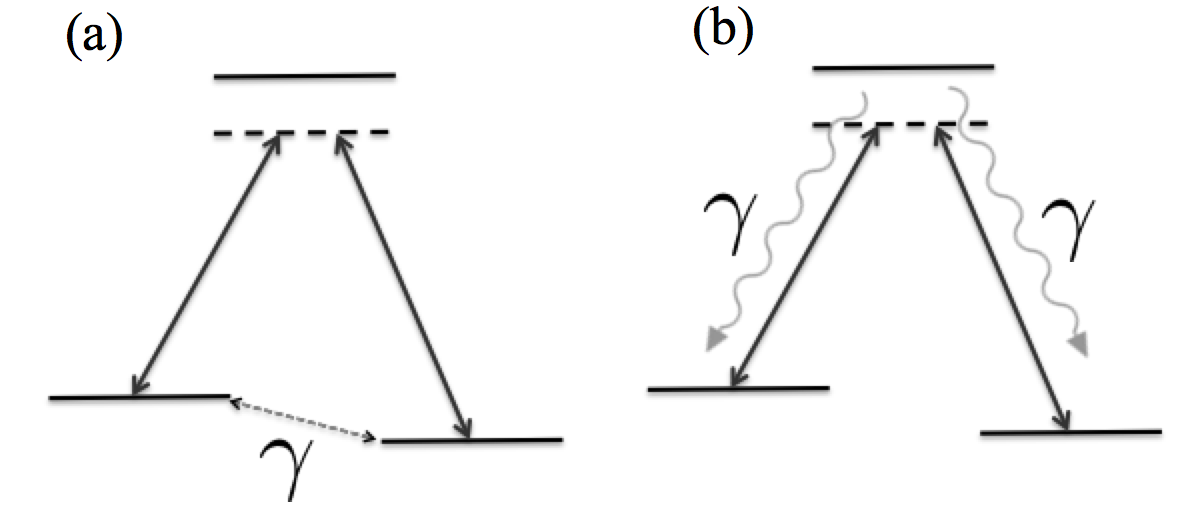}}
\caption{Schematic picture of the two open STIRAP situations: (a) dephasing of the two stable $|1\rangle$ and $|3\rangle$ states, and (b) spontaneous emission from the excited $|2\rangle$ state to either of the lower states. } \label{fig3}
\end{figure} 

The system evolution is modeled within the Lindblad master equation formalism, Eq.~(\ref{master}). In deriving Eq.~(\ref{master}), the {\it Markov}, {\it Born}, and {\it secular} approximations have been employed which are typically justified in the optical regime of interest here~\cite{open}. Errors stemming from the explicit time-dependence of the Hamiltonian combined with using an {\it ad hoc} application of the Lindblad master equation is assumed to give no qualitative changes of the conclusions~\cite{stirapmaster}. We will work in the diabatic basis such that the Hamiltonian is taken as $\hat H_\mathrm{d}$ of Eq.~(\ref{stirapham}), and we consider two different situations of loss channels:
\begin{enumerate}

\item {\it Case} ($a$). Dephasing of the lower states $|1\rangle$ and $|3\rangle$ implemented by the Lindblad jump operator
\begin{equation}\label{jump1}
\hat L=|1\rangle\langle 1|-|3\rangle\langle 3|.
\end{equation}
We disregard any dephasing between the other levels as these are typically of less importance~\cite{dephase1}, and we are more interested in a qualitative understanding of the underlying physics and not on a qualitative description. 

\item {\it Case} ($b$). Spontaneous emission of the excited state $|2\rangle$ to the states $|1\rangle$ and $|3\rangle$. The corresponding jump operators are
\begin{equation}\label{jump2}
\hat L_1=|1\rangle\langle 2|,\hspace{0.7cm}\hat L_2=|3\rangle\langle2|.
\end{equation}
Note that we do not study the situation of losses to a fourth level $|4\rangle$ as was the scenario of Refs.~\cite{sponem1,sponem2}. Furthermore, we assume the same decay rates, $\gamma_1=\gamma_2=\gamma$.

\end{enumerate}
These two cases are schematically presented in Fig.~\ref{fig3}. 

Before expressing the Lindblad master equation~(\ref{master}) in the Bloch representation, we give some remarks already captured in the traditional representation of the master equations. In general it might not be relevant to look at steady states for a time-dependent Hamiltonian, but if, however, the loss rate $\gamma$ would set the fast time-scale it is interesting to ask the question what the `instantaneous' steady state solutions $\partial_t\hat\rho_\mathrm{ss}=0$ will be. Hence, we consider fixed $G_1(t)$ and $G_2(t)$ and solve for the steady state $\mathcal{L}[\hat\rho]=0$. Without losses, the adiabatic states~(\ref{adstate}) are the only steady states; $\hat\rho_\mathrm{ss}=|\varphi_i(t)\rangle\langle\varphi_i(t)|$ with $i=0,\,\pm$. For case ($b$) with jump operators~(\ref{jump2}) we find that the dark state $|\varphi_0(t)\rangle$ is also an eigenstate of these jump operators with zero eigenvalue, {\it i.e.} $\hat L_i|\varphi_0(t)\rangle=0$ ($i=1,\,2$). This, then implies that the `instantaneous' steady state for this case is $\hat\rho_\mathrm{ss}=|\phi_0(t)\rangle\langle\varphi_0(t)|$. The Lindblad jump operator (\ref{jump1}) for case ($a$) is diagonal in the diabatic basis, and thus any diagonal state which commutes with $\hat H_\mathrm{d}(t)$ is a steady state. Indeed, in this case the maximally mixed state $\hat\rho_\mathrm{ss}=\mathbb{1}/3$, with $\mathbb{1}$ the identity operator, clearly fulfils this and is thereby the instantaneous steady state. As a consequence, if the influence of the environment is too strong we can directly conclude that the STIRAP transfer rate will be detrained.

\subsection{Some general properties of the master equation and its Liuvillian} \label{subsec:M}
We may rewrite any Lindblad master equation~(\ref{master}) as a set of coupled first order differential equations for the matrix elements $\rho_{ij}=\langle i|\hat\rho|j\rangle$ for some basis states $|i\rangle$. By reshuffling the elements to form a vector with length $D^2$ ($D$ being the dimension of the problem), we obtain an ordinary differential equation written on a matrix form. However, here we will not consider such a parametrization, but rather use the generalized Bloch vector as our parametrization. 

Any state can be written on the form~\cite{genpara} 
\begin{equation}
\hat\rho=\frac{1}{D}\left(\mathbb{1}+\sqrt{\frac{D(D-1)}{2}}\,\mathbf{R}\cdot\lambda\right),
\end{equation}
where $\mathbf{R}=(r_1,\,r_2,\,\dots,\,r_{D^2-1})$ is the generalized {\it Bloch vector} and $\lambda=(\lambda_1,\,\lambda_2,\,\dots,\,\lambda_{D^2-1})$ a vector with the {\it generalized Gell-Mann matrices} as elements~\cite{ggm}. These $\lambda_i$ are the generators of $SU(D)$; for $D=2$ (qubits) they become the Pauli matrices, while for $D=3$ (qutrits) they are instead the standard Gell-Mann matrices (see Appendix~\ref{sec:app2}). The generalized Gell-Mann matrices are orthogonal in any dimension, {\it i.e.} $\mathrm{Tr}[\lambda_i\lambda_j]=2\delta_{ij}$, such that given $\hat\rho$ the Bloch vector elements $r_i$ are easily obtainable by multiplying with the correct matrix $\lambda_i$ and then perform the trace. For pure states one has $|\mathbf{R}|=R=1$, and naturally for the maximally mixed state $\mathbf{R}=\mathbf{0}$. Solving the master equation for $\hat\rho$ now transforms into the problem of solving an equation on the form~\cite{openad,openad2} 
\begin{equation}\label{blochev}
\partial_t\mathbf{R}=\mathbf{M}\mathbf{R}+\mathbf{b}.
\end{equation}
Here $\mathbf{M}$ is a $(D^2-1)\times(D^2-1)$ matrix generating the time evolution of the Bloch vector, and will be hereafter called the {\it Liouvillian matrix}, while $\mathbf{b}$ we denote {\it Liouvillian pump} for reasons to be explained below. Note first that $\mathbf{M}$ is uniquely determined once the Hamiltonian and the Lindblad jump operators are specified, and secondly that for $\mathbf{b}\neq0$, Eq.~(\ref{blochev}) is non-homogeneous. One may, as in Refs.~\cite{openad,openad2}, define new vectors $\tilde{\mathbf{R}}=(1,\,r_1,\,r_2,\dots,\,r_{D^2-1})$ and $\tilde\lambda=(\mathbb{1},\,\lambda_1,\,\lambda_2,\,\dots,\,\lambda_{D^2-1})$ and the corresponding equation for $\tilde{\mathbf{R}}$ will always be homogeneous. Here, however, we stick to the more standard representation of the density operator in terms of a Bloch vector. Written on the form~(\ref{blochev}) the physical meaning of the $\mathbf{b}$ term also becomes more transparent.

The Liouvillian matrix $\mathbf{M}$ is normally not hermitian, meaning that its eigenvalues may be complex and furthermore one has to separate between left and right eigenvectors. Even though the left and right eigenvectors may be different, there always exists a similarity matrix $\mathbf{S}$ such that
\begin{equation}\label{jordan}
\mathbf{D}=\mathbf{S}\mathbf{M}\mathbf{S}^{-1},
\end{equation}
where the matrix $\mathbf{D}$ is on {\it Jordan-Block form}~\cite{jform}. That is, given one of the blocks it is comprised of degenerate eigenvalues on its diagonal and ones on the superdiagonal. Blocks of dimension higher than one typically appear at the EP's. For any hermitian matrix $\mathbf{H}(\chi)$, where $\chi$ is any set or real-valued parameters, the EP's are found in the complex plane of the $\chi$'s by analytical continuation. Here, however, when we have a non-hermitian matrix the EP's may lie along the real $\chi$-axes, for example if $\mathbf{M}$ is time-dependent along the real time axis. We note that in the special situation of a closed system, the eigenvectors of $\mathbf{M}$ corresponds to the states $\hat\rho_{ij}=|i\rangle\langle j|$ where $|i\rangle$ is the $i$'th eigenstate of the Hamiltonian. Another property of the Liouvillian matrix is that it is real and thereby its trace is also real, which has the consequence that its eigenvalues can be grouped in pairs with opposite imaginary parts. 

There is a further important observation to be stressed about the Bloch representation of the master equation. For any physical state $\hat\rho$ we must have that it is: ($i$) hermitian, ($ii$) positive semidefinite, and ($iii$) normalized with unit trace. For qubits, {\it i.e.} $D=2$, all states with $0\leq R\leq1$ are physical, implying that the Bloch sphere comprises all the allowed states. In higher dimensions, qitrits and more generally qudits, only part of the Bloch sphere represents actual physical states~\cite{genpara,genpara2}. We mentioned above that the eigenstates of the master equation for a closed system are $\hat\rho_{ij}=|i\rangle\langle j|$, meaning that there are $D$ times as many eigenstates for the Liouvillian matrix $\mathbf{M}$ than to the Hamiltonian. These additional states are, of course, those with $i\neq j$ which are traceless and hence do not represent physical states. Note further that $\hat\rho_{ij}=|i\rangle\langle j|$ gives $D^2$ states, but the dimension of $\mathbf{M}$ is $D^2-1$. This extra state, missing in our Bloch representation, appears since the $\hat\rho_{ii}$ states are not linearly independent (we have $\sum_{i=1}^D\hat\rho_{ii}=\mathbb{1}$) and one of them is always redundant. For non-zero coupling to the environment one typically finds most states unphysical, and interestingly the number of physical states may not be the same as for the closed case. A direct requirement for having a physical state is that the corresponding eigenvalue of $\mathbf{M}$ is real (otherwise $\mathbf{R}$ is complex and $\hat\rho$ not hermitian). We have found for our three-level problem that whenever we have a real $\mathbf{R}$ the corresponding state $\hat\rho$ is physical provided that $R$ is sufficiently small, {\it i.e.} there is a ball surrounding the origin $\mathbf{R}=\mathbf{0}$ containing only physical states. As a result, for the open STIRAP problem we need a minimum of two real eigenvalues~\footnote{It can be shown that the real part of an eigenvalue of $\mathbf{M}$ has to be nonpositive~\cite{x}.} of $\mathbf{M}$ (remember that they are grouped in pairs) and the corresponding states will be physical. As we will see in the next section, for large couplings $\gamma$ it is indeed possible to have all states physical (and thereby all eigenvalues real). It is intriguing that the unphysical states may show interesting properties that are missing for the physical ones. In Ref.~\cite{geo} the geometric phase of these were studied and it was shown that this phase is normally non-vanishing, while for the physical states the geometric phase strictly vanishes. Geometric phases connected to encircling EP's have been explored as well~\cite{epgeo}, but the link between the phases found for non-physical states and those related to EP's were not studied in Ref.~\cite{geo}.

While unphysical, these states are still crucial for the time evolution~\cite{decay}. If $\mathbf{b}=\mathbf{0}$, and we denote the right eigenvectors $\mathbf{R}^{(i)}$ we have that the time evolved Bloch vector can be expressed
\begin{equation}\label{te}
\mathbf{R}(t)=\sum_{i=1}^8c_ie^{\mu_it}\mathbf{R}^{(i)},
\end{equation}
where $\mu_i$ are the eigenvalues of the Liouvillian, $c_i$ are the coefficients determined by the initial state $\mathbf{R}(0)$, and we have assumed a time-independent Liouvillian matrix $\mathbf{M}$. Even though the right (left) eigenvectors are not orthogonal in general, the series expansion is unique given that the vectors are normalized and we do not `sit' on an EP. The sum will in general include unphysical Bloch vectors, which does not imply that the full sum $\mathbf{R}(t)$ cannot be physical. Indeed, it has to be as long as $\mathbf{R}(0)$ is physical. Note that we must have $\mathrm{Re}(\mu_i)\leq0$~\cite{x}. If $\mu_i=0$, the corresponding Bloch vector $\mathbf{R}^{(i)}$ is stationary and if the state $\hat\rho^{(i)}$ is physical it is consequently a steady state of the master equation. Whenever the eigenvalues contain real parts, these will cause an exponential decay of the corresponding terms in the sum. If the Lindblad jump operators $\hat{L}_i$ are hermitian, the maximally mixed state $\hat\rho=\mathbb{1}/D$ is a steady state and if this is also the unique steady state we must have that all eigenvalues have non-zero real parts~\cite{spohn}. The eigenvalue with the smallest amplitude of its real part, {\it i.e.}
\begin{equation}\label{LG}
\tilde\Delta=\min_i\left[\mathrm{Re}(-\mu_i)\right],
\end{equation}
sets an upper bound for the time-scale for reaching the steady state, and it is often referred to as the {\it Liouvillian gap}~\cite{decay,decay2}. 

For $\mathbf{b}\neq\mathbf{0}$, which typically occurs for non-hermetian Lindblad jump operators $\hat L_i$, the situation is more complicated. Here, a stationary state is represented by a Bloch vector $\mathbf{R}=-\mathbf{M}^{-1}\mathbf{b}$ given that $\mathbf{M}$ is invertible. If $\mathbf{M}$ is not invertible, however, the system is underdetermined and you find a connected manifold of solutions. Introducing the matrices $\mathbf{V}$ and $\mathbf{U}$ such that $\mathbf{E}=\mathbf{V}^t\mathbf{MU}$ with $\mathbf{E}$ diagonal, the right eigenvectors of $\mathbf{M}$ evolve in time as
\begin{equation}
\mathbf{R}^{(i)}(t)=\mathbf{R}^{(i)}(0)e^{\mu_it}+\left(e^{\mu_it}-1\right)\mathbf{V}^t\mathbf{b}/\mu_i.
\end{equation}
For $\mathbf{b}=\mathbf{0}$ we recover the exponential time-dependence of Eq.~(\ref{te}). We see, as stated above, that for $\mathbf{b}\neq\mathbf{0}$ a negative real part of $\mu_i$ is not sufficient to warrant that the steady state is the simple maximally mixed state as for hermitian Lindblad jump operators. In these situations, looking at the dynamical equation (\ref{blochev}) it is clear that the $\mathbf{b}$ term acts as a sort of `pump' which has the effect of restoring some coherence/purity of the state $\hat\rho(t)$. 

It might well happen that $\mathbf{M}$ is not be diagonalizable. This happens at the so called {\it exceptional points} in the parameter space~\cite{expo}. At an EP, the real parts of two eigenvalues, $\varepsilon_1(\nu)$ and $\varepsilon_2(\nu)$ (with $\nu$ some parameter), coalesce and at this `degeneracy point' the corresponding eigenstates become identical, in stark contrast to a degeneracy of an Hermitian operator were the eigenstates can always be chosen orthogonal. The EP is also characterized by a square-root singularity~\cite{x}, and upon encircling such a singularity once, the corresponding eigenstates are interchanged, while encircling it twice gives you back the same state up to an overall sign change~\cite{epgeo,expo2}. This generalizes the geometric phase appearing when enclosing a {\it Dirac point} in condensed matter physics or a {\it conical intersection} in molecular/chemical physics~\cite{bohm} to degeneracy points for non-hermitian operators. In the following section we will see that for the present model, numerous EP's appear in the Liouvillian matrices as the parameter $\gamma$ is varied keeping the remaining parameters fixed.

\section{Results and discussions}\label{sec3}
In this section we numerically solve the master equation (\ref{blochev}) for the two cases represented by the jump operators (\ref{jump1}) and (\ref{jump2}). Building on what we concluded in the previous section about the Liouvillian matrix $\mathbf{M}$ we analyze our numerical findings. Before doing so, however, we need to resolve what the properties of $\mathbf{M}$ are for the open STIRAP problems.

\subsection{The Liouvillian matrix for STIRAP}
The Liouvillian matrices, together with some of their properties, are presented in the Appendix~\ref{sec:app1} for the two cases~(\ref{jump1}) and (\ref{jump2}). As the couplings $G_1(t)$ and $G_2(t)$ now are time-dependent different scales become relevant. We will come back to this in the following subsection for the general case, while here we fix the values of the couplings. In particular we consider $G_1=G_2=1$ and $\Delta=0$. We have verified that varying these particular values does not alter our conclusions. With these parameters fixed we ask how the eigenvalues of $\mathbf{M}$ varies with the system-environment coupling $\gamma$. Naturally, keeping the couplings constant gives an oversimplified picture (loosely speaking assuming that the short time-scale is that of environment induced relaxation), but it do provide valuable insight.

We may note from the previous discussion, for $\gamma=0$ and since the Liouvillian matrices are skew-symmetric in this limit we find two zero eigenvalues. From the expression (\ref{M1}) it is easy to identify the dark state Bloch vector
\begin{equation}\label{darkbloch}
\mathbf{R}_0=\left[
\begin{array}{cccccccc}
0 & 0 & \sqrt{3}c^2 & -\sqrt{3}sc & 0 & 0 & 0 & (1-3s^2)
\end{array}\right]^t/2
\end{equation}
as one of them, with $s=\sin\theta$ and $c=\cos\theta$ defined below Eq.~(\ref{adstate}). As $\gamma$ becomes non-zero the two zero eigenvalues split but stay real (and negative)~\cite{x}. The dark state Bloch vector is no longer a zero eigenvalue eigenstate which is seen from the non-zero matrix element $\mathbf{M}_{44}=-2\gamma$. Nevertheless, as pointed out above, the dark state is still a stationary state for case ($b$) which is only possible since the pump vector $\mathbf{b}$ is non-zero. This is somewhat unintuitive since one usually think about a non-zero Liouvillian gap as driving the system to a maximally mixed state.

\begin{figure}[h]
\centerline{\includegraphics[width=8cm]{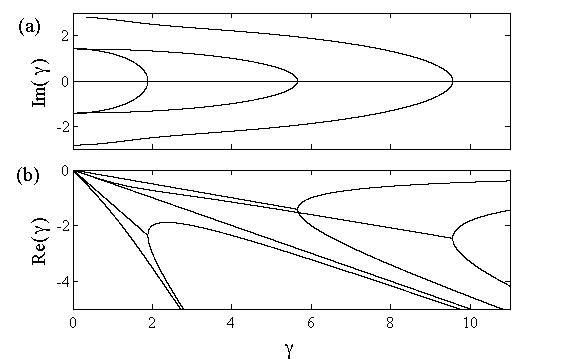}}
\caption{The imaginary (a) and real (b) parts of the eigenvalues for case ($a$); dephasing of the lower states $|1\rangle$ and $|3\rangle$. As long as $\gamma\neq0$ all eight eigenvalues possess a negative real part implying that the Bloch vector decays to the origin. The EP's occur when the imaginary part vanishes (happening in pairs) and the real parts display a bifurcation-like bahaviour. The dimensionless parameters are $G_1=G_2=1$ and $\Delta=0$.} \label{fig4}
\end{figure} 

\begin{figure}[h]
\centerline{\includegraphics[width=8cm]{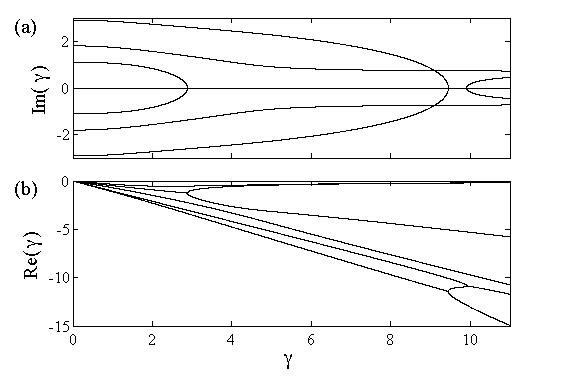}}
\caption{Same as Fig.~\ref{fig4} but for case ($b$); spontaneous emission of the excited state $|2\rangle$. A similar structure is found as for case ($a$), but with one new feature, namely the appearance of non-zero imaginary parts (seen around $\gamma\approx10$). Even though the Liouvillian gap $\tilde\Delta\neq0$ whenever $\gamma\neq0$ it is possible to find a non-trivial steady state thanks to the non-vanishing pump term $\mathbf{b}$ (see the Appendix~\ref{sec:app1}). Note that both in this figure and in Fig.~\ref{fig4} we consider rather large values of $\gamma$ in order to demonstrate the general properties. } \label{fig5}
\end{figure}

In Figs.~\ref{fig4} and \ref{fig5} we display the real and imaginary parts of the eigenvalues of the Liouvillian matrices for case ($a$) and ($b$) respectively. In both cases we have that the Liouvillian gap $\tilde\Delta$ is non-zero whenever the coupling to the environment is present. For case ($a$), where $\mathbf{b}=\mathbf{0}$, this results in that the steady state is the maximally mixed state with $\mathbf{R}=\mathbf{0}$. For case ($b$) the real parts of the eigenvalues are also all non-positive, but the pump makes it possible that the steady state contains quantum coherences as we have already argued. The grouping of the eigenvalues is evident, the imaginary parts occur in $\pm$ pairs. The disappearance of imaginary parts at the EP's implies a splitting of the real parts, reminiscent of a bifurcation. At these points the eigenvectors become real, but in order for them to represent physical states their lengths must be relatively short. Indeed, as already mentioned, given that the Bloch vector is real it is always possible to construct a physical state $\hat\rho$ from it given that one shrinks its length sufficiently much~\cite{genpara2}. In case ($a$), for large enough $\gamma$'s all eigenstates become purely real after the three EP's. In case ($b$), however, a `reversed' EP takes place where two purely real eigenstates become imaginary upon increasing $\gamma$.

\subsection{Dynamics -- numerical results}
\begin{figure}[h]
\centerline{\includegraphics[width=8cm]{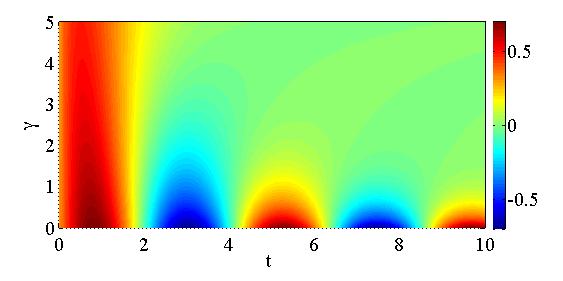}}
\caption{Population imbalance between the two lower states for case ($a$) as a function of time and system-environment coupling $\gamma$ for an initial random pure state. For small $\gamma$ the evolution at this scale is predominantly unitary with clear Rabi oscillations between the two states, while for larger $\gamma$ the decay of the imbalance is approximately exponential. The parameters are as in the previous figures, $G_1=G_2=1$ and $\Delta=0$. } \label{fig6}
\end{figure}

The previous Subsection analyzed general effects stemming from the environment, and not the interplay between them and the inherent STIRAP dynamics. This is the main focus of this Subsection, but before so we consider one more scenario with the couplings $G_1$ and $G_2$ constant. A relevant question is whether the non-analyticity characteristic for the EP's of the eigenvalues encountered in Fig.~\ref{fig4} can also lead to non-analytic evolution of any initial state. The connection between EP's and non-equilibrium phase transitions has been discussed in the past~\cite{EPPT}, and one may expect signatures of these also in the present setting. For closed systems, {\it i.e.} Hamiltonian with a real spectrum, non-analyticity in the spectrum may also arise but in the thermodynamic limit, and as for the EP's they can give rise to {\it excited state phase transitions} which may naturally affect the evolution~\cite{ept}. In particular, one could expect that the time-evolution of some initial state will qualitatively change if one varies $\gamma$; whenever a splitting of the real parts in Fig.~\ref{fig4} occurs, the exponents rendering the decay change which could alter the system evolution. We explore this by calculating the population imbalance $Z=\mathrm{Tr}\left[(|1\rangle\langle1|-|3\rangle\langle3|)\hat\rho(t)\right]$ for an initial random pure state. The result for case ($a$) is shown in Fig.~\ref{fig6}. For small $\gamma$ ($<G_1,\,G_2$) and sufficiently short times, the evolution is dominated by unitary time evolution, {\it i.e.} the population imbalance displays Rabi oscillations between the two states. When the decay is increased, it determines the fast time-scale and the relaxation occurs before any Rabi oscillations take place. The presence of non-analytic features of the eigenvalues are, however, not reflected in the time evolution. Thus, for random initial states we do not encounter something like excited state phase transitions (we have tried for numerous different initial random states). The situation might change, though, if the initial state coincide with a single eigenstate of the Liouvillian matrix, and the change in $\gamma$ is adiabatic in the sense of Ref.~\cite{openad}.

If the Liouvillian gap $\tilde\Delta$ is large in comparison to the inverse time that sets the inherent time evolution, adiabaticy is not guaranteed by a large adiabaticity parameter $a$. Generally speaking, a large parameter $a$ favours internal adiabatic evolution, but it implies an extended coupling to the environment which in return tends to take the system out of its instantaneous adiabatic eigenstate~\cite{openad,openad2}. Consequently, there should be an optimal $a_{opt}$ such that the intrinsic unitary evolution is close to adiabatic and at the same time excitations due to the environment are not too definite. This scenario should apply to case ($a$) representing dephasing, while for case ($b$) there should not be a trade-off between the two processes, {\it i.e.} a slow passage does not automatically lead to environment induced excitations since the state remains in a dark state. Nevertheless, as we will see, in the quasi-adiabatic regime when the intermediate state becomes slightly populated the environment will affect the STIRAP efficiency.

\begin{figure}[h]
\centerline{\includegraphics[width=8cm]{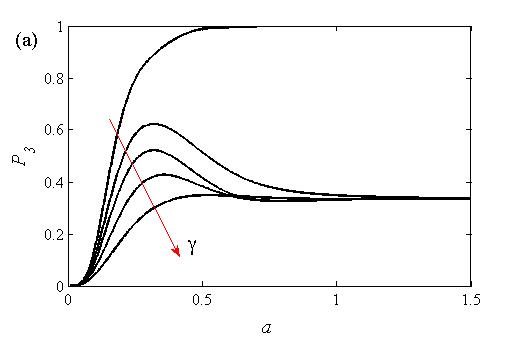}}
\centerline{\includegraphics[width=8cm]{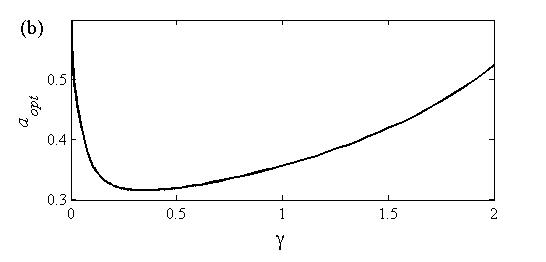}}
\caption{Upper plot (a): Final population of the target state $|3\rangle$ in the case ($a$) for different adiabaticity parameters $a$ and five different loss rates $\gamma=0,\,1/4,\,1/2,\,1,\,2$ in growing order with the arrow. In the closed case, upper curve for $\gamma=0$, the population is a monotonously increasing function of $a$ demonstrating that adiabaticity is increased with a large $a$. As explained in the main text, for an open STIRAP process, the lower curves, there is an optimal $a_{opt}$ which maximises the population transfer. The lower plot (b) displays this optimal adiabaticity parameter as a function of the loss rate $\gamma$. The remaining dimensionless parameters are $g_0=1$, $\Delta=0$, and $\tau=\sigma=10$. } \label{fig7}
\end{figure}

The numerical results for the full time-dependent STIRAP problem for case ($a$) are displayed in Fig.~\ref{fig7}. As a measure of the efficiency of STIRAP we calculate the final population of the state $|3\rangle$, {\it i.e.} $P_3$ introduced in the previous Section. The integration interval $t\in[t_i,t_f]$ ($t_f=-t_i=100$) is chosen long enough such that convergence has been reached. As we know from Fig.~\ref{fig2} (b), for the closed STIRAP $a$ serves as an adiabaticity parameter, {\it i.e.} there is a one-to-one relation between its value and the success rate of the STIRAP. For the cases of Fig.~\ref{fig7} ($\gamma$ ranges from 0 to 2), at sufficiently large $a$ ($>0.3$) the steady state $
\hat\rho_{ss}=\mathbb{1}/3$ is reached as expected from the discussion above after Eq.~(\ref{jump2}).  The appearance of this maximally mixed state has also been numerically verified by calculating the purity $P=\mathrm{Tr}\left[\hat\rho^2\right]$ (not shown here). The $\gamma$-dependence of the optimal $a_{opt}$ is shown in Fig.~\ref{fig7} (b). For larger values of $\gamma$ beyond 2 there is no longer a clear maximum any longer. In the limit $\gamma\rightarrow0$ we know that $a_{opt}$ diverges which is not clear from the figure since we actually use a finite integration interval. In the parameter regime where $P_3$ builds up a maximum as a function of $a$ coincide with the scenario when $a\tau\sim\gamma^{-1}$, {\it i.e.} there is no clear separation between time-scales of the inherent STIRAP process and the external decay rate.   

\begin{figure}[h]
\centerline{\includegraphics[width=8cm]{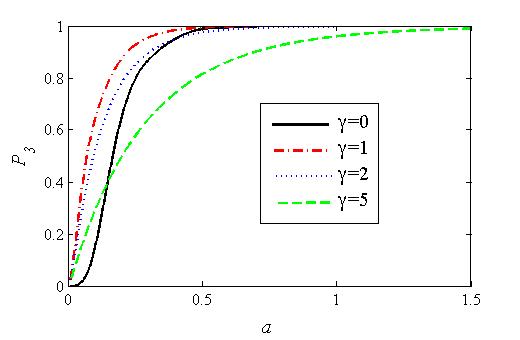}}
\caption{Same as the previous Fig.~\ref{fig7} but for case ($b$) describing spontaneous emission of the intermediate $|2\rangle$ level. As was argued in the main text, since the dark state is an instantaneous steady state for slow processes, {\it i.e.} large $a$'s, the system state is transparent to the environment. For fast processes when adiabaticity breaks down it is found that the environment actually increases the population transfer, which can be subscribed the non-vanishing Liouvillian gap. However, there is a tradeoff, a too strong coupling to the bath may lower the success rate as seen from the green curve. Indeed, in the limit of a very large $\gamma$ we recover the Zeno effect of Ref.~\cite{stirapmaster} which manifest as the system frozen in the initial state $|1\rangle$.} \label{fig8}
\end{figure} 

As mentioned already, case ($b$) is conceptually different from case ($a$). The STIRAP dark state $|\varphi_0(t)\rangle$ is a zero eigenvalue eigenstate of the Lindblad jump operators $\hat L_{1,2}$ of Eq.~(\ref{jump2}), meaning that it is transparent to the environment. Perfect population transfer is thereby expected in the deep adiabatic regime. This is also found numerically, as demonstrated in Fig.~\ref{fig8}. Contrary to the case of dephasing, here the population of the target state is always a monotonously increasing function of $a$ regardless of the value of $\gamma$. Or put in other words, $a$ is a proper parameter to characterize adiabaticity in the process. However, it is not fully clear how adiabaticity should be defined or characterized in open quantum systems. This issue has been raised in a series of papers~\cite{openad,openad2,openad3}. One definition is to define it by saying that there should be no population transfer between the different Jordan blocks~(\ref{jordan}). This definition is natural as the Jordan blocks for open systems play the same role as degenerate subspaces for closed systems. Following this, the adiabatic criterium~(\ref{adcrit}) can be modified to open systems in a rather direct way~\cite{openad}. As is clear, the openness may greatly affect the criteria and typically the environment induces additional excitations~\cite{opencrit}. In this respect, something interesting occurs for small $a$ when the STIRAP is not fully adiabatic; the spontaneous emission increases the efficiency of the population transfer. This derives from the presence of a Liouvillian gap $\tilde\Delta$ which implies that the dark state is partly protected from non-adiabatic excitations. So contrary to the standard situation where the fluctuations from the environment induces excitations, in the non-adiabatic regime the environment prevents the system from taking it out from its instantaneous state. This phenomenon is related, but still different from the Zeno-effect discussed in Ref.~\cite{stirapmaster}. There, in the overdamped regime, the environment prohibited the system from leaving the initial state $|1\rangle$. In the present situation we are far from overdamped and the environment induced relaxation takes you back onto the dark adiabatic state and not onto the initial state. 

The loss mechanism~(\ref{jump2}) describes the decay into an incoherent mixture of $|1\rangle$ and $|3\rangle$. One may ask whether a `coherent' decay~\cite{diehl}, like for example represented by the Lindblad jump operator $\hat L=|1\rangle\langle2|+|3\rangle\langle2|$, would affect the result of Fig.~\ref{fig8}. We tried for a couple of different such jump operators and found that the coherence does not play a quantitative role, any such decay protects the adiabatic state for fast processes. 

\section{Conclusion}\label{sec4} 
Due to the importance of the STIRAP mechanism, many studies on the open STIRAP situation have been presented in the past. In the present work we took, however, an unusual approach to the problem, and in such a way provided new insight in the underlying mechanisms. In particular, we expressed the Lindblad master equation in terms of generalized Bloch equations. The resulting Liouvillian matrix $\mathbf{M}$, rendering the time evolution, was studied in detail for the most relevant scenarios; dephasing and spontaneous emission. The matrix $\mathbf{M}$ has a complex spectrum which, to a large extent, explains the system properties. For example, real negative parts of the eigenvalues cause an exponential decay towards some steady state, and non-analyticities of the spectrum occur in terms of EP's. In the case of dephasing this instantaneous steady state is the maximally mixed state which would mean that the STIRAP application totally fails as long as the system-environment coupling is strong enough ({\it i.e.} sets the shortest time-scale). As a result, there is an optimal process time for the population time: the system generated dynamics is not too fast (non-adiabatic) while the decay into the mixed state is not complete. The situation is, however, very different for spontaneous emission of the excited atomic state. Here the dark state is prone to dissipation and a complete STIRAP population transfer is indeed possible even for a strong system-bath coupling. While this is a known result, in terms of the Liouvillian formulation one understands this from the interplay between a decay (negative real parts of the eigenvalues of $\mathrm{M}$) and a non-zero `pump term' ($\mathbf{b}\neq0$). In particular, the pump restores coherence/purity of the system state. While it would be interesting to explore in more detail what kind of Lindblad jump operators resulting in different system evolution, it seems that for Hermitian operators the pump term $\mathbf{b}$ is always absent. Physically this is not strange since such operators typically describe decohering processes while non-Hermitian operators give dissipation of any sort. 

We further found that the environment could even increase the population transfer. This happens when the STIRAP is not perfect to start with, {\it i.e.} not perfectly adiabatic. In this case the non-zero Liouvillian gap has the effect of taking your state back onto the dark state. Thus, the environment assists the transfer. This is, however, not true for all parameters; as the transfer becomes more complete the environment does no longer aid the transition.

In addition to analyzing the influence of the environment on the STIRAP process, we also discussed the mathematical structure of the Liouvillian matrix in some general terms. In particular, the dimension of $\mathbf{M}$ is greater than that of the system Hamiltonian, meaning in general that not all eigenstates represent physical states. As a result, these states may possess properties which one would not find in those states that are physical. Even though some of the states are unphysical they do, however, play an important role in the system evolution since sums of unphysical eigenstates can add up to a physical total state.

\appendix

\section{Gell-Mann matrices}\label{sec:app2}
For the three-level STIRAP problem, rewriting the density operator on the Bloch form implies expanding it in the identity plus the regular Gell-Mann matrices~\cite{georgi} 
\begin{equation}\label{gm}
\begin{array}{lll}
\lambda_1=\left[\begin{array}{ccc}
0 & 1 & 0\\
1 & 0 & 0\\
0 & 0 & 0
\end{array}\right], & & \lambda_2=\left[\begin{array}{ccc}
0 & -i & 0\\
i & 0 & 0\\
0 & 0 & 0
\end{array}\right],\\ \\ 
\lambda_3=\left[\begin{array}{ccc}
1 & 0 & 0\\
0 & -1 & 0\\
0 & 0 & 0
\end{array}\right], & & \lambda_4=\left[\begin{array}{ccc}
0 & 0 & 1\\
0 & 0 & 0\\
1 & 0 & 0
\end{array}\right],\\ \\ 
\lambda_5=\left[\begin{array}{ccc}
0 & 0 & -i\\
0 & 0 & 0\\
i & 0 & 0
\end{array}\right], & & \lambda_6=\left[\begin{array}{ccc}
0 & 0 & 0\\
0 & 0 & 1\\
0 & 1 & 0
\end{array}\right],\\ \\ 
\lambda_7=\left[\begin{array}{ccc}
0 & 0 & 0\\
0 & 0 & -i\\
0 & i & 0
\end{array}\right], & & \lambda_8=\frac{1}{\sqrt{3}}\left[\begin{array}{ccc}
1 & 0 & 0\\
0 & 1 & 0\\
0 & 0 & -2
\end{array}\right],\\ \\ 
\end{array}
\end{equation}
The Gell-Mann matrices (as their generalizations to higher dimensions) are orthogonal in the sense that $\mathrm{Tr}\left[\lambda_i\lambda_j\right]=2\delta_{ij}$, and furthermore they are traceless. The commutation relations are
\begin{equation}\label{gm2}
\left[\lambda_i,\lambda_j\right]=i2f^{ijk}\lambda_k,
\end{equation}
with the antisymmetric tensor $f^{ijk}$ obeying
\begin{equation}
\begin{array}{c}
f^{123}=1,\\ \\
f^{147}=f^{165}=f^{246}=f^{257}=f^{345}=f^{376}=\displaystyle{\frac{1}{2}},\\ \\
f^{458}=f^{678}=\displaystyle{\frac{\sqrt{3}}{2}}.
\end{array}
\end{equation}
Any remaining elements, for example with two or three identical superscripts equal, are identically zero. 

\section{STIRAP Liouvillian matrices}\label{sec:app1}
Employing the commutation properties of the Gell-Mann matrices~(\ref{gm2}) the derivation of the Liouvillian matrix and pump is straightforward. In the case ($a$) representing pure dephasing (see Eq.~(\ref{jump1})) the pump term vanishes as argued in Subsec.~\ref{subsec:M}. The matrix is found to be
\begin{equation}\label{M1}
\mathbf{M}\!=\!\!\left[\!\!
\begin{array}{cccccccc}
\!-\gamma\!/2 & \Delta & 0 & 0 & G_2 & 0 & 0 & 0\\
-\Delta & \!-\gamma\!/2 & \!-2G_1 & \!-G_2 & 0 & 0 & 0 & 0\\
0 & 2G_1 & 0 & 0 & 0 & 0 & -G_2 & 0\\
0 & G_2 & 0 & -2\gamma & 0 & 0 & -G_2 & 0\\
-G_2 & 0 & 0 & 0 & -2\gamma & G_1 & 0 & 0\\
0 & 0 & 0 & 0 & \!-G_1 & \!-\gamma\!/2 & -\Delta & 0\\
0 & 0 & G_2 & G_1 & 0 & \Delta & -\gamma\!/2 & \!-\sqrt{3}G_2\\
0 & 0 & 0 & 0 & 0 & 0 & \!\sqrt{3}G_2 & 0
\end{array}\!\!\!\right]\!\!.
\end{equation}
Note that $\mathbf{M}$ is skew-symmetric and real. In odd dimensions the skew-symmetric property implies that the matrix is singular, but this is not the case here as the dimension is eight. The skew-symmetry property also implies that the eigenvalues are paired as pointed out in the main text. Another observation is that the number of elements proportional to $G_2$ is higher than those proportional to $G_1$ which might seem spurious at first. The reason for this is in the definition of the Gell-Mann matrices~(\ref{gm}); there is an asymmetry between the different $SU(2)$ subgroups of the $SU(3)$ group and the atomic transitions are connected to two different subgroups.

For the second case ($b$) of a dissipating excited state $|2\rangle$ (see Eq.~(\ref{jump2})), the jump operators are no longer hermitian and we indeed find a nonzero pump contribution,
\begin{equation}\label{M2}
\mathbf{M}\!=\!\!\left[\!\!
\begin{array}{cccccccc}
\!-\gamma\!/2 & \Delta & 0 & 0 & G_2 & 0 & 0 & 0\\
-\Delta & \!-\gamma\!/2 & \!-2G_1 & -G_2 & 0 & 0 & 0 & 0\\
0 & 2G_1 & -\gamma & 0 & 0 & 0 & -G_2 & \gamma\!/\!\sqrt{3}\\
0 & G_2 & 0 & -\gamma\!/2 & 0 & 0 & -G_2 & 0\\
-G_2 & 0 & 0 & 0 & \!-\gamma\!/2 & G_1 & 0 & 0\\
0 & 0 & 0 & 0 & -G_1 & \!-\gamma & -\Delta & 0\\
0 & 0 & G_2 & G_1 & 0 & \Delta & -\gamma & \!-\sqrt{3}G_2\\
0 & 0 & 0 & 0 & 0 & 0 & \!\sqrt{3}G_2 & 0
\end{array}\!\!\!\right]\!\!,
\end{equation}
and
\begin{equation}
\mathbf{b}^t=\left[
\begin{array}{cccccccc}
0 & 0 & \gamma/\sqrt{3} & 0 & 0 & 0 & 0 & 0
\end{array}\right]
\end{equation}
Contrary to the first case, the matrix is no longer skew-symmetric due to a non-zero term $\mathbf{M}_{38}=\gamma/\sqrt{3}$. It is clear that this contribution is connected to the non-vanishing pump $\mathbf{b}$ which has also a non-zero term as its third element. 

\begin{acknowledgements}
The authors thank M. Sarandy for helpfull discussions, and acknowledge financial support from  KAW (The Knut and Alice Wallenberg foundation) and VR-Vetenskapsr\aa set (The Swedish Research Council).
\end{acknowledgements}


\begin{thebibliography}{999}

\bibitem{shore} B. W. Shore, {\it Manipulating Quantum Structures Using Laser Pulses}, (Cambridge University Press, Cambridge 2011).

\bibitem{cc} Y. Nakamura, Yu. A. Pashkin, and J. S. Tsai, Nature {\bf 398}, 786 (1999); N. S. Ginsberg, S. R. Garner, and L. Vestergaard Hau, Nature {\bf 445}, 623 (2006); M. Shapiro and P. Brumer, {\it Principles of the quantum control of molecular processes}, (Wiley, 2012).

\bibitem{cclz} H.-J. Miesner, D. M. Stamper-Kurn, J. Stenger, S. Inouye, A. P. Chikkatur, and W. Ketterle, Phys. Rev. Lett. {\bf 82}, 2228 (1999); L. Gaudreau, G. Granger, A. Kam, G. C. Aers, S. A. Studenikin, P. Zawadzki, M. Pioro-Ladrière, Z. R. Wasilewski	, A. S. Sachrajda, Nature Phys. {\bf 8}, 54 (2012). 

\bibitem{vitanovrev} N. V Vitanov, T. Halfmann, B. W Shore, and K. Bergmann, Ann. Phys. Rev. Phys. Chem. {\bf 52}, 763 (2001); N. V. Vitanov, A. A. Rangelov, B. W. Shore, and K. Bergmann, arXiv:1605.00224.

\bibitem{sponem1} N. V. Vitanov and S. Stenholm, Phys. Rev. A {\bf 56}, 1463 (1997).

\bibitem{sponem2} P. A. Ivanov, N. V. Vitanov, nd K. Bermann, Phys. Rev. A {\bf 72}, 053412 (2005).

\bibitem{dephase1} P. A. Ivanov, N. V. Vitanov, and K. Bergmann, Phys. Rev. A {\bf 70}, 063409 (2004).

\bibitem{open} H. P. Breuer and F. Petruccione, {\it The theory of pen quantum systems}, (Oxford University Press, Oxford 2002).

\bibitem{expo} W. D. Heiss, J. Phys. A: Math. Theor. {\bf 37}, 2455 (2004); W. D. Heiss, J. Phys. A: Math. Theor. {\bf 45}, 444016 (2012). 

\bibitem{openad} M. S. Sarandy and D. A. Lidar, Phys. Rev. A {\bf 71}, 012331 (2005).

\bibitem{openad2} M. S. Sarandy and D. A. Lidar, Phys. Rev. Lett. {\bf 95}, 250503 (2005).  

\bibitem{BOA} M. Bear, {\it Beyond Born-Oppenheimer}, (Wiley, 2006).

\bibitem{adtheo} A. Messiah, {\it Quantum Mechanics}, (North-Holland publishing company, 1961).

\bibitem{stirapmaster} M. Scala, B. Militello, A. Messina, N. V. Vitanov, Phys. Rev. A {\bf 81}, 053847 (2010).

\bibitem{strong}  F. Beaudoin, J. M. Gambetta, and A. Blais, Phys. Rev. A {\bf 84}, 043832 (2011).

\bibitem{be} F. Bloch, Phys. Rev. {\bf 70}, 460 (1946).

\bibitem{genpara} G. Kimura, Phys. Lett. A {\bf 314}, 339 (2003); R. A. Bertlmann and P. Krammer, J. Phys. A: Math.Theor. {\bf 41}, 235303 (2008).

\bibitem{ggm} F. T. Hioe, J. H. Eberly, Phys. Rev. Lett. {\bf 47}, 838 (1981).

\bibitem{jform} R. A. Horn and C. R. Johnson, {\it Matrix Analysis}, (Cambridge University Press, Cambridge, 1999). 

\bibitem{genpara2} S. K. Goyal, B. N. Simon, R. Singh, and S. Simon, J. Phys. A: Math. Theor. {\bf 49} 165203 (2016).
 
\bibitem{geo} S. Dasgupta and D. A. Lidar,  J. Phys. B {\bf 40}, S127 (2007). 

\bibitem{epgeo} A. A. Mailybaev, O. N. Kirillov, and A. P. Seyranian, Phys. Rev. A {\bf 72}, 014104 (2005).

\bibitem{decay} M. V. Medvedyeva and S. Kehrein, Phys. Rev. B {\bf 90}, 205410 (2014).

\bibitem{x} E. M. Kessler, G. Giedke, A. Imamoglu, S. F. Yelin, M. D. Lukin, and J. I. Cirac, Phys. Rev. A {\bf 86}, 012116 (2012).

\bibitem{spohn} H. Spohn, Lett. Math. Phys. {\bf 2}, 33 (1977).

\bibitem{decay2} Z.  Cai  and  T.  Barthel,  Phys.  Rev.  Lett. {\bf 111},  150403 (2013).

\bibitem{expo2} H. Mehri-Dehnavi and A. Mostafazadeh, J. Math. Phys. {\bf 49}, 082105 (2008).

\bibitem{bohm}  A. Bohm, A. Mostafazadeh, H. Koizumi, Q. Niu, and J. Zwanziger, {\it The Geometric Phase in Quantum Systems}, (Springer Verlag, Berlin, 2003).

\bibitem{EPPT} W. D. Heiss, M. M\"uller, and I. Rotter, Phys. Rev. E {\bf 58}, 2894 (1998); I. Rotter, J. Phys. A: Math. Theor. {\bf 42}, 153001 (2009). 

\bibitem{ept} M. A. Caprio, P. Cejnar, F. Iachello, Ann. Phys. {\bf 323}, 1106 (2008); T. Brandes, Phys. Rev. E {\bf 88}, 032133 (2013).

\bibitem{openad3} P. Thunstr\"om, J. \AA berg, and E. Sj\"oqvist, Phys. Rev. A {\bf 72}, 022328 (2005); L.C. Venuti, T. Albash, D. A. Lidar, and P. Zanardi, Phys. Rev. A {\bf 93}, 032118 (2016).

\bibitem{opencrit} J. Dziarmaga, Phys. Rev. B {\bf 74}, 064416 (2006); J. Dziarmaga, Adv. in Phys. {\bf 59}, 1063 (2010).

\bibitem{diehl} S. Diehl, A. Micheli, A. Kantian, B. Kraus, H. P. B\"uchler, and P. Zoller, Nature Phys. {\bf 4}, 878 (2008).

\bibitem{georgi} H. Georgi, {\it Lie Algebras in Particle Physics}, (Boulder, Westview Press, 1999).

\end{thebibliography}
\end{document}